\documentclass[a4paper,fleqn]{cas-dc}

\usepackage[numbers]{natbib}
\usepackage{tabularx}
\newcommand{\rev}[1]{{\color{black}#1}}
\newcommand{\new}[1]{{\color{black}#1}}
\usepackage{booktabs}
\usepackage{array}
\usepackage{makecell}

\newcolumntype{L}{>{\raggedright\arraybackslash}X}
\newcolumntype{C}{>{\centering\arraybackslash}X}
\def\tsc#1{\csdef{#1}{\textsc{\lowercase{#1}}\xspace}}
\tsc{WGM}
\tsc{QE}
\tsc{EP}
\tsc{PMS}
\tsc{BEC}
\tsc{DE}

\begin{document}
\let\WriteBookmarks\relax
\def\floatpagepagefraction{1}
\def\textpagefraction{.001}
\shorttitle{RISC-V Empowered Blockchain}
\shortauthors{Qinglin~Yang et~al.}

\title [mode = title]{Embedded Blockchain Infrastructure Management (eBIM): A RISC-V-Empowered Hardware--Software Co-Design Framework Towards Trustworthy Blockchain}                      

\tnotetext[1]{This document is the result of the research project funded by the Smart Grid-National Science and Technology Major Project Grant 2025ZD0805900; in part by the GuangDong Basic and Applied Basic Research Foundation under Grant 2025B1515020022, 2026A1515010183; in part by the National Natural Science Foundation of China under Grant T2522011.}

\author[1]{Qinglin Yang}[ orcid=0000-0002-7263-8914]



\credit{Conceptualization of this study, Methodology, Software}

\affiliation[1]{organization={Cyberspace Institute of Advanced Technology, Guangzhou University, Guangdong Key Laboratory of Industrial Control System Security, and Huangpu Research School of Guangzhou University},
                city={Guangzhou},
                postcode={510006}, 
                state={Guangdong},
                country={China}}

\author[1]{Yuan~Liu}
\cormark[1]
\ead{{yuanliu}@gzhu.edu.cn}
\author[1]{Yaoyao Zhang}
\author[1]{Boya Wang}
\author[1]{Zongjian You}

\credit{Data curation, Writing - Original draft preparation}

\author[2]{Chunming Rong}
\cormark[1]
\ead{chunming.rong@uis.no}

\author[1]{Zhihong Tian}
[style=chinese]

\affiliation[2]{organization={Department of Computer and Electrical Engineering, University of Stavanger},
                country={Norway}}

\cortext[cor1]{Corresponding author}


\begin{abstract}
Blockchain systems are undergoing a fundamental transition from decentralized ledgers for digital assets to general-purpose trust infrastructures for verifiable computation, decentralized physical resources, and automated infrastructure management. \rev{However, deploying these systems through third-party Blockchain-as-a-Service (BaaS) offerings may introduce dependencies involving operational control, confidentiality, and infrastructure governance. Meanwhile, blockchain execution and verification impose demanding requirements on cryptographic processing, isolation, and reproducibility. RISC-V offers an open and extensible instruction set architecture on which software execution environments and domain-specific security hardware can be co-designed.
Inspired by the embedded identity and provisioning model of an embedded subscriber identity module (eSIM), this survey proposes embedded Blockchain infrastructure management (eBIM) as a conceptual hardware–software co-design framework that organizes existing RISC-V-enabled technologies into a hardware-anchored blockchain infrastructure stack. 
We systematically review RISC-V-based smart-contract execution, trusted execution environments, zero-knowledge virtual machines, and cryptographic acceleration, and analyze their relationship to blockchain requirements, including deterministic execution, verifiability, extensibility, and secure key management.
Based on this evidence, we distinguish currently demonstrated capabilities from the proposed eBIM integration vision and identify major open challenges, including system-level benchmarking, blockchain-oriented cryptographic extensions, TEE security, post-quantum migration, and deployable hardware–software co-design. The survey provides an architectural reference for researchers and designers investigating open and verifiable blockchain infrastructure.}
%
%

\end{abstract}

\begin{graphicalabstract}
\includegraphics[width=0.6\textwidth]{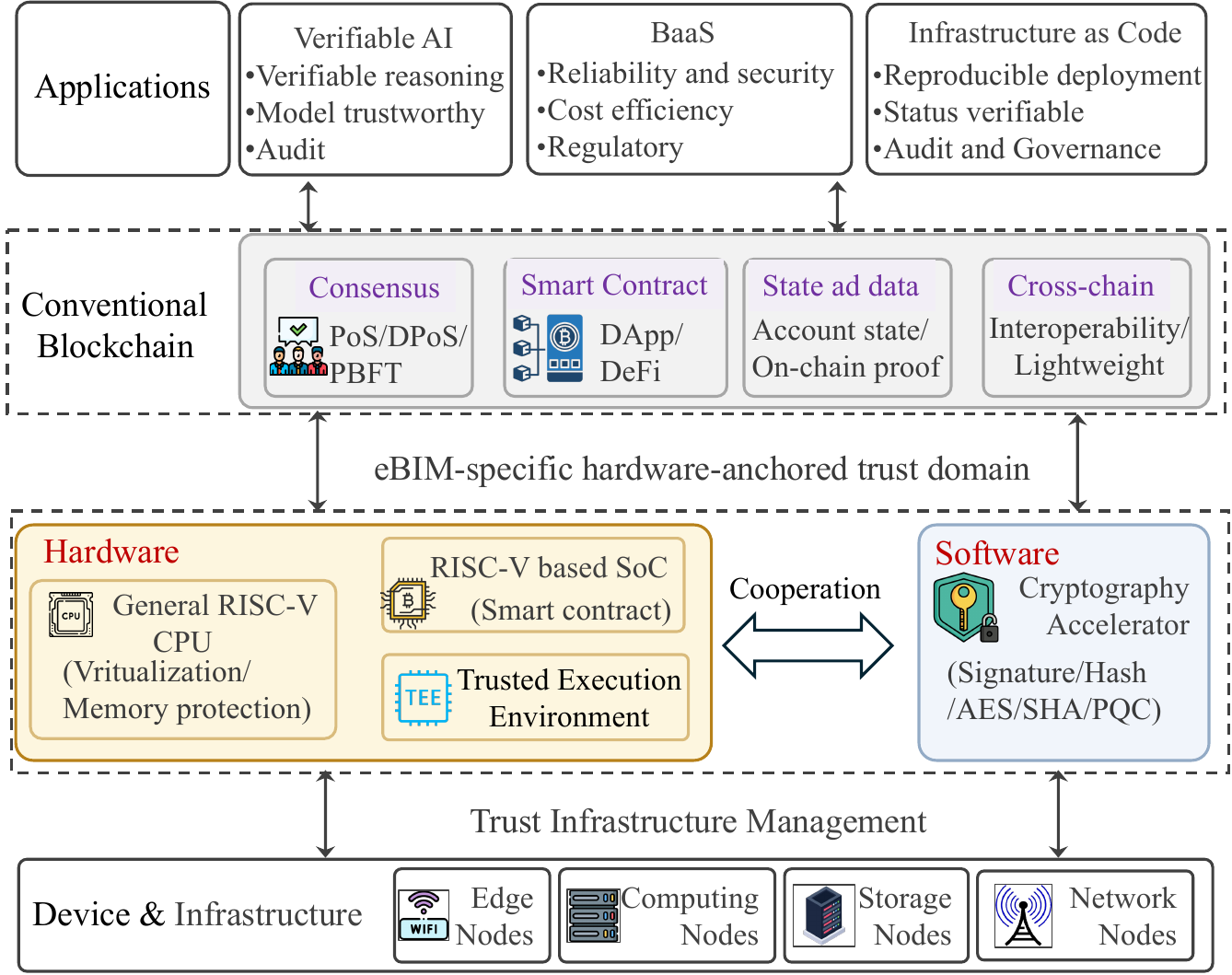}
\end{graphicalabstract}

\begin{highlights}
\item It defines eBIM as a conceptual hardware–software co-design framework and specifies its trust boundaries, capabilities, and relationship to adjacent technologies.

    \item It maps representative literature on RISC-V contract execution, cryptographic acceleration, TEEs, and zkVMs to the framework's components and workflow. It distinguishes established results from architectural proposals.
  \item This paper identifies key challenges and future research directions, including blockchain-oriented RISC-V cryptographic extensions, post-quantum security, and system-level hardware-software co-design for eBIM. 
\end{highlights}


\maketitle
  \sloppy
\section{Introduction}\label{sec:Introduction}

The prevailing discourse in blockchain research has shifted decisively from theoretical justification toward practical scalability.
Specifically, how distributed ledger systems can be responsibly scaled to meet the demands of real-world, high-throughput deployments. In this context, data privacy is not an ancillary concern but a foundational architectural requirement. Financial institutions operating on blockchain infrastructure are subject to stringent regulatory frameworks governing the confidentiality of client data and transactional activity. Similarly, public-sector deployments must reconcile the competing imperatives of citizen data protection and systemic auditability. For digital assets, tokenized financial instruments, and verifiable data infrastructures to achieve mainstream institutional adoption, underlying blockchain platforms must provide robust, scalable confidentiality guarantees.

While blockchain systems employ consensus mechanisms to establish agreement among distributed nodes and maintain consistent replicated states, their execution environments, particularly blockchain virtual machines (VMs), continue to face significant challenges in computational efficiency, verifiable computation~\cite{chaliasos2024snarksok}, cryptographic security, and long-term maintainability~\cite{chu2023survey,zhang2025smartupdater,xiang2025bytecodecomment}.

The European Blockchain Service Infrastructure (EBSI) and the China Blockchain Service Network (BSN) represent two main paths in the development of global sovereign blockchains, which are cloud-based service model that enables organizations to build, deploy, and operate blockchain applications without designing, configuring, or maintaining the underlying blockchain infrastructure. 
Despite their different governance philosophies, they face a highly similar engineering challenge: how to provide a trusted, manageable, and interoperable identity and authentication system for the massive number of nodes, devices, and users connected to the infrastructure?

Concurrently, the instruction set architecture (ISA), \new{reduced instruction set computing (RISC-V)~\cite{waterman2014risc}}, has emerged as a transformative paradigm in processor design.
Distinct from proprietary architectures such as x86, dominated by Intel and AMD, and ARM, RISC-V’s open, modular, and extensible design makes it a uniquely promising foundation for blockchain and verifiable-computation systems~\cite{tain2025survey}. 
This modular architecture enables several complementary capabilities, such as targeted optimization of cryptographic primitives, integration of domain-specific hardware accelerators, and adaptation to heterogeneous execution environments.
%

%
%
%

From a blockchain system architecture perspective, recent studies have explored hardware acceleration strategies for smart contract execution to mitigate the computational overhead inherent to virtual-machine-based execution models.
Collectively, these convergent developments establish RISC-V as a credible architectural pathway toward efficient, formally verifiable, and scalable blockchain system design \cite{lu2023scu}. For example, Polkadot's deployment of PolkaVM on its Westend testnet~\cite{polkavm2025}. Nervos CKB's long-standing use of CKB-VM as its RISC-V contract execution layer\cite{Xue2019CKBVM}. Meanwhile, the emergence of RISC-V-based zkVM projects SP1 and RISC Zero as leading zero-knowledge proving platforms~\cite{succinct2025sp1}. For blockchain and verifiable-computation systems, RISC-V provides an open and extensible execution substrate that supports smart-contract execution, serves as the instruction-set foundation for zero-knowledge virtual machines (zkVMs), and enables the implementation of TEEs for isolating security-sensitive workloads~\cite{Open2026Application}.

An embedded subscriber identity module (eSIM), as used in cellular systems, supports device identity and remotely managed subscriber profiles within a controlled security and provisioning ecosystem. 
Inspired by eSIM's capabilities~\cite{zhao2021securesim} in hardware-level security isolation and standardized identity authentication, the trusted framework~\textit{embedded blockchain infrastructure management (eBIM)} is proposed, which is a software-hardware collaborative blockchain infrastructure management model.
We treat eSIM as an architectural analogy, not as evidence that blockchain nodes can inherit eSIM security guarantees. In the proposed eBIM framework, an embedded trust anchor could protect node-specific credentials, support attestation, and expose well-defined interfaces to transaction execution and verification services. 

At its core, eBIM leverages RISC-V as the foundation for designing customized secure hardware chips, bringing the cryptographic acceleration, key management, and trusted execution of critical contract logic required for blockchain operations down to the hardware layer. \rev{eBIM is intended to provide autonomous and verifiable blockchain infrastructure capabilities} for nodes and edge devices, without reliance on centralized cloud service providers. It differs from hardware security module (HSM)-based approaches in that eBIM embeds programmable, extensible contract execution logic within the hardware boundary itself, rather than delegating contract logic to an external software layer.

\subsection{Motivations and Contributions}

Despite the rapid development of RISC-V-based blockchain applications in both academia and industry, the existing literature remains fragmented and lacks a systematic survey that explains how RISC-V empowers blockchain from the perspectives of architecture, execution, cryptography, verification, and applications. It is critically important to answer the following three questions (RQs) comprehensively around eBIM:
\begin{enumerate}
   \item \textbf{RQ1} What is eBIM?
   This research question aims to provide the composition of eBIM, including what this novel paradigm entails, how its multi-layered architecture is structured, and what distinguishes it from prior art. 
   \item \textbf{RQ2} How does eBIM work?
   This research question aims to systematically reveal how eBIM anchors the entire lifecycle of blockchain transactions within the RISC-V hardware boundary, thereby translating decentralized security requirements into concrete, actionable hardware-enforced guarantees. 
   \item \textbf{RQ3} What can eBIM do? 
   This research question aims to define the core capabilities that eBIM delivers to blockchain nodes and hardware-embedded trust,  transforming blockchain infrastructure from an external service into embedded intrinsic trust.

\end{enumerate}

Our contributions are summarized as follows:

\begin{enumerate}
\item It defines eBIM as a conceptual hardware–software co-design framework and specifies its trust boundaries, capabilities, and relationship to adjacent technologies.

    \item It maps representative literature on RISC-V contract execution, cryptographic acceleration, TEEs, and zkVMs to the framework's components and workflow. It distinguishes established results from architectural proposals.
  \item This paper identifies key challenges and future research directions, including blockchain-oriented RISC-V cryptographic extensions, post-quantum security, and system-level h
\end{enumerate}

\begin{figure}
    \centering
    \includegraphics[width=1\linewidth]{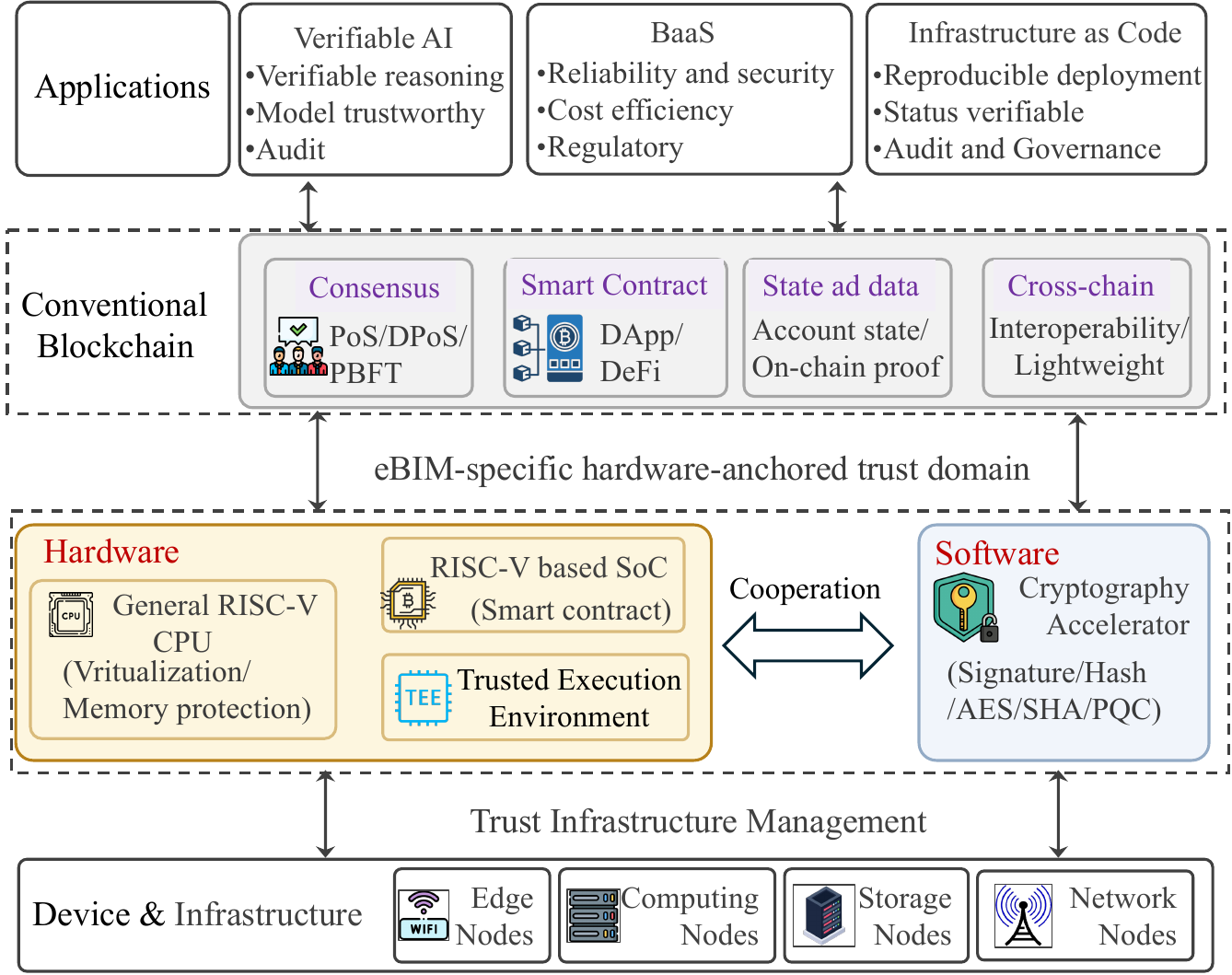}
    \caption{The illustration of the hierarchical architecture of eBIM.}
    \label{fig:illustration}
\end{figure}
The rest of this work is organized as follows. Section~\ref{sec:Background} provides the fundamental introduction to RISC-V and blockchain; Section~\ref{sec:feature_comatching} describes how RISC-V features empower blockchain; Section~\ref{sec:Performance} introduces the performance of RISC-V from two aspects: software execution and hardware infrastructure; Section~\ref{sec:Application} introduces the main application of RISC-V; Section~\ref{sec:Challenges} provides the future research direction of RISC-V; finally, Section~\ref{sec:Conclusions} concludes this paper.

\section{Background and Related work}\label{sec:Background}

This section briefly provides the basic concepts, such as \rev{RISC-V}, blockchain, and smart contracts. Furthermore, this section reviews the most related work in the fields of blockchain and RISC-V.

\subsection{The Features of RISC-V}

The key properties of RISC-V are most clearly reflected in its modularity, open standard, and verification-friendly design, which is evident in its typical execution flow~\cite{riscv-unpriv-2024,riscv-ratified-specs}.
Instruction set architectures are commonly classified into two broad categories: i) complex instruction set computing (CISC) and ii) reduced instruction set computing (RISC).

Classic RISC/CISC studies \cite{hennessy2019new,celio2016renewed} have shown that CISC tends to reduce the number of program-level instructions by providing a richer and more complex instruction repertoire. 
RISC-V adopts a modular ''base instruction set \& standard extensions" paradigm, which allows users to select and combine features according to specific application requirements~\cite{riscv-unpriv-2024,riscv-ratified-specs}.

For example, in software development, programmers typically write code in high-level languages such as C/C++ and then rely on a RISC-V compiler toolchain to translate source code into machine code executable on the target processor. Existing GNU RISC-V toolchains and GCC already support RISC-V target platforms ~\cite{riscv-gnu-toolchain,gcc-riscv-options}.
At the microarchitectural level, many classical or instructional RISC-V-based processor designs adopt a five-stage pipeline consisting of instruction fetch (IF), instruction decode (ID), execution (EX), memory access (MEM), and write back (WB), allowing multiple instructions to progress concurrently through different stages and thereby improving instruction throughput. Nevertheless, the five-stage pipeline should be understood as one common implementation style rather than a universal template for all RISC-V processors \cite{chang2024five-stage,jin2025five-stage}. When interrupts or exceptions occur, the RISC-V privileged architecture can flexibly route traps to handlers at the appropriate privilege level and resume execution through the corresponding trap-return instruction, thereby supporting external event handling and fault recovery \cite{riscv-priv-html-alt}. 

Therefore, RISC-V's manifest advantages are summarized as follows: i) Reducing interpreter overhead; ii) Reducing control-flow jumps for loop unrolling issues; iii) Using smaller data types instead of defaulting to 256-bit words as in the Ethereum virtual machine (EVM)~\cite{buterin2016ethereum}. These advantages show that the performance improvement of RISC-V does not come entirely from the instruction set itself, but rather from a more direct and lower-overhead execution model.

\subsection{Blockchain and Smart Contract}

\begin{figure*}
    \centering
    \includegraphics[width=0.9\linewidth]{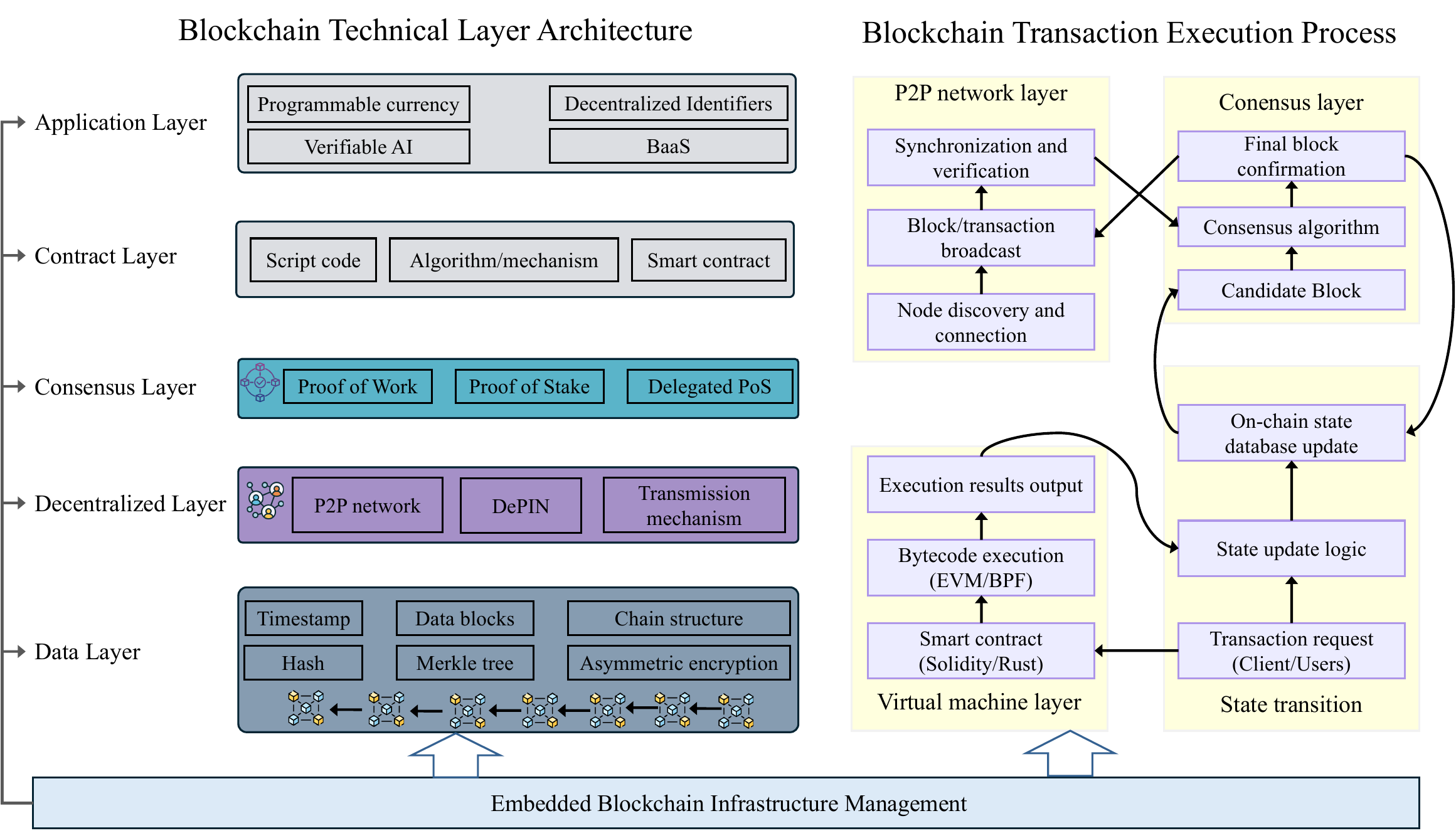}
    \caption{The layer architecture of blockchain technology and the main execution process of blockchain transactions, including the virtual machine, state transition, consensus, and decentralized layer.}
    \label{fig:blockchain}
\end{figure*}

Blockchain technology~\cite{zheng2017overview} is a collection of concepts and technologies that form a distributed ledger maintained by a peer-to-peer network in which transactions are grouped into cryptographically linked blocks. The key technologies that underpin blockchain systems include cryptography \& hash functions, block structure, consensus mechanisms, and smart contracts, as demonstrated in Fig.\ref{fig:blockchain}.

Hash functions are deterministic algorithms (such as SHA-256) that convert input data of any size into a fixed-length output. Any change to the input produces a completely different hash, making them ideal for detecting tampering and ensuring data integrity. 
In blockchains, hashes link blocks together and verify transaction authenticity. Public key cryptography (e.g., ECDSA, RSA, EdDSA) enables asymmetric encryption where each user has a public key and a private key, which enables participants to prove ownership of assets and authorize transactions without exposing their private keys. Digital signatures combine hashing and public-key cryptography to create unforgeable proof of identity and authorization.

Smart contracts~\cite{khan2021blockchain} are self-executing programs deployed on blockchain networks that automatically enforce contractual agreements through deterministic code execution. Formally, a smart contract is a stateful computational entity resident on the blockchain whose behavior is entirely determined by its bytecode, input parameters, and the current state of the distributed ledger. Once deployed to a specific blockchain address, a smart contract's code becomes immutable and accessible to all network participants, enabling transparent, auditable, and trustless contract execution. Smart contracts are typically written in high-level languages such as Solidity (for Ethereum), Move (for Aptos/Sui), or Rust (for Solana and other systems), which are subsequently compiled to low-level bytecode interpretable by the blockchain's virtual machine execution environment.

\subsection{Comparisons with the Existing Studies}

Existing trusted hardware platforms like Intel SGX and MIT Sanctum lack formal security guarantees. Developers and hardware architects could not rigorously reason about whether enclave platforms truly protected against privileged software adversaries, making it impossible to systematically compare platforms or verify that security properties held under different threat models.

To this end, Cerdeira~\emph{ et al.}~\cite{cerdeira2020sok} provide the first comprehensive cross-vendor quantitative analysis of trusted execution environment (TEE) security, establishing a lower bound on vulnerability prevalence. Subramanyan~\emph{et al.}~\cite{subramanyan2017formal} introduce the trusted abstract platform (TAP), a formal idealization of enclave execution defined as a finite state transition system. They decomposed the high-level property of secure remote execution (SRE) into three formally verifiable sub-properties: integrity, confidentiality, and secure measurement. Then, they developed machine-checked proofs in BoogiePL/Z3 showing that formal models of Intel SGX and MIT Sanctum are refinements of the TAP under specific adversary parameterizations, meaning every execution trace of SGX or Sanctum can be mapped to a corresponding TAP trace. Javaid~\emph{et al.}~\cite{javaid2022blockchain} propose Blockchain machine (BMac), a hardware/software co-designed network-attached accelerator implemented on a Xilinx Alveo U250 FPGA. The architecture includes two primary components: a protocol processor and a block processor. The protocol processor implements a custom UDP-based protocol that breaks blocks into self-contained packets, replaces repetitive X.509 identity certificates with compact 16-bit encoded IDs (removing at least 73\% of block size), and annotates packets with field offsets enabling cut-through hardware parsing without TCP-style full-block buffering. 
The block processor includes block-level and transaction-level pipelines. The 2-stage block-level pipeline processes multiple blocks in a pipelined fashion. In contrast, an internal 3-stage transaction pipeline (tx\_verify, tx\_vscc, tx\_mvcc\_commit) validates multiple transactions in parallel using configurable numbers of ECDSA engine instances, endorsement policy evaluation via combinational circuits with short-circuit logic, and an in-hardware key-value state database implemented in BRAM/URAM.

To improve the performance of permissioned blockchains that apply FPGA-based hardware accelerators, Agrawal~\emph{et al.}~\cite{agrawal2022efficient} propose several optimizations for modular arithmetic and point arithmetic. At the modular arithmetic level, they developed a hybrid schoolbook-Karatsuba integer multiplier that completes 256-bit multiplication in 39 clock cycles by combining symmetric 32-bit word decomposition with Karatsuba's technique on 16-bit sub-words, efficiently utilizing FPGA DSP blocks. For modular reduction over P-256, they proposed an incremental correction version of the NIST fast Mersenne reduction that avoids wide parallel comparators by checking and correcting after each addition/subtraction step.

In summary, the aforementioned studies address complementary layers of the same trust infrastructure stack. They first provide the formal theoretical foundation establishing what security properties an enclave platform must satisfy and how to verify them. Then, they address the performance layer, indicating how FPGA-based hardware/software co-design can overcome the cryptographic bottlenecks that limit blockchain throughput, precisely the hardware acceleration core of eBIM.

\section{The Description of eBIM}
\subsection{Security assumptions and attack surface of eBIM.}
\new{Hardware anchoring does not eliminate trust. Instead, it relocates and makes part of the trusted computing base explicit. The eBIM model assumes that the hardware root of trust, secure boot chain, attestation keys, and security monitor are correctly provisioned and protected. An adversary may target the chip supply chain, firmware, shared microarchitectural resources, external memory interfaces, secure-key storage, cryptographic accelerators, or remote-attestation infrastructure. Physical probing, fault injection, power/electromagnetic leakage, and cache-timing attacks are particularly relevant to edge devices exposed to potentially hostile environments. Therefore, an eBIM implementation requires complementary mechanisms such as secure boot, authenticated firmware update, access control on the SoC interconnect, memory isolation, constant-time cryptographic implementation, side-channel-resistant key storage, and attestation-key lifecycle management.}

Accordingly, ``hardware-anchored trust” in eBIM should be understood as reducing and localizing selected trust assumptions rather than claiming absolute hardware security.

\subsection{Architecture details of eBIM}

The proposed eBIM is a reference architecture with three distinct execution boundaries. (1) On-chip trusted services comprise secure key storage, signing or verification engines, and potentially an isolated execution environment. (2) Node software runs the blockchain client, policy enforcement, networking, and any proof-system orchestration on a host processor or operating system. (3) The decentralized network performs peer-to-peer propagation, consensus, and state replication across independent nodes. A RISC-V system-on-chip (SoC) can integrate a TEE-capable execution subsystem. Still, an SoC is not itself synonymous with a TEE: isolation depends on implemented privilege separation, memory protection, trusted boot, attestation, and other security mechanisms. Fig.~\ref{fig:illustration} presents a trusted blockchain infrastructure framework that demonstrates how physical devices, trusted execution hardware/software, blockchain services, and upper-layer applications connect.

At the bottom layer, the system starts with Device \& Infrastructure, including edge nodes, computing nodes, storage nodes, and network nodes. These nodes provide the physical foundation and distributed resources for the whole framework.

The next layer is divided into Hardware and Software, which work in cooperation. The hardware side includes a general RISC-V CPU for virtualization and memory protection, a RISC-V-based SoC for smart contract support, and a TEE. The software side includes a cryptography accelerator that supports operations such as signature, hash, AES, SHA, and post-quantum cryptography (PQC).

Above the hardware and software is the Execution \& Validation layer. This layer connects trusted computing resources with blockchain functions. It ensures that blockchain operations can be executed, verified, and validated in a trusted way.

The Blockchain layer sits above execution and validation. It contains four major components: consensus such as PoS, DPoS, and PBFT; smart contracts for DApp and DeFi logic; state and data for account state and on-chain proof; and cross-chain mechanisms for interoperability and lightweight access.

At the top is the Applications layer. It includes three representative application scenarios: Verifiable AI, supporting verifiable reasoning, model trustworthiness, and audit; BaaS, emphasizing reliability, security, cost efficiency, and regulation; and Infrastructure as Code, supporting reproducible deployment, verifiable status, audit, and governance.

\new{eBIM lies in the systematic hardware–software integration of blockchain execution, cryptographic operations, key management, and trusted execution around an open and extensible RISC-V substrate. The novelty of eBIM is not a replacement of the conventional blockchain protocol stack, but the introduction of a programmable hardware-anchored trust substrate underneath and across the execution/validation boundary.}

\begin{figure}
    \centering
    \includegraphics[width=0.49\textwidth]{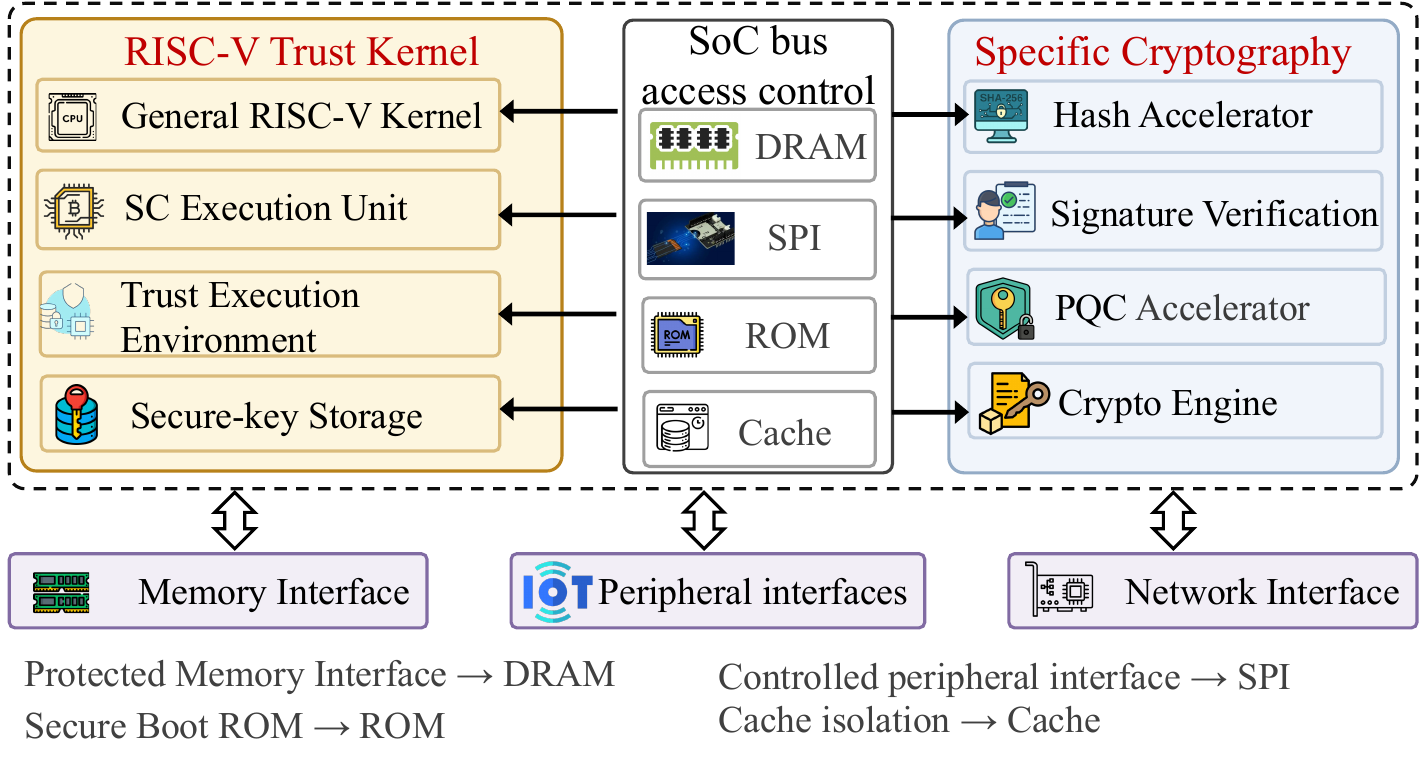}
    \caption{\rev{The module architecture illustration of the proposed eBIM}.}
    \label{fig:EBIMChip}
\end{figure}
\textbf{Answer to RQ1.} Embedded blockchain infrastructure management is a software-hardware collaborative blockchain infrastructure management paradigm. At its core, eBIM leverages the open instruction set architecture (RISC-V) as the foundation for designing customized secure hardware chips, bringing the cryptographic acceleration, key management, and trusted execution of critical contract logic required for blockchain operations down to the hardware layer, as depicted in Fig~\ref{fig:EBIMChip}. \new{Specifically, ROM is associated with the immutable or protected boot root used to establish the initial trust chain. DRAM access is mediated through memory protection and isolation mechanisms. Cache is identified as a security-sensitive shared microarchitectural resource requiring isolation or side-channel mitigation. SPI represents an external peripheral interface whose accesses should be controlled and authenticated when security-critical devices are attached.}

\section{Why eBIM is Potential?}\label{sec:feature_comatching}
RISC-V’s verification-friendly characteristics are particularly evident in the decode, execute, and write-back stages, where regular instruction formats, well-defined operational semantics, and explicit state-update mechanisms facilitate systematic verification.

The relevance of RISC-V to blockchain applications stems from its open instruction set architecture (ISA), established compiler ecosystem, and support for domain-specific extensions. These characteristics provide a foundation for auditable implementations and specialized hardware acceleration. However, they do not inherently guarantee system-wide determinism, security, or reduced latency. Therefore, this section examines the relationship between RISC-V and blockchain by analyzing how RISC-V’s architectural characteristics align with key blockchain requirements, including transparency, deterministic execution, cryptographic security, and verifiable state transitions. 


\subsection{Openness and transparency}\label{sub:Openness}
Blockchain aims to reduce dependence on centralized entities through public rules, verifiable ledgers, and consensus mechanisms, which require not only transparent protocols but also auditable execution environments.
Furthermore, blockchain consensus requires different nodes to produce the same output from the same transaction and state input. The execution environment must therefore avoid uncontrolled nondeterministic behavior. 
RISC-V has clear instruction boundaries, explicit register operations, well-defined memory access rules, and structured state updates. RISC-V's properties make the execution behavior of blockchain easier to constrain and analyze.

\subsection{Scalability of Blockchain}

From the perspective of blockchain scalability, the value of RISC-V is not that it simply replaces the EVM, but that it reduces the additional overhead caused by EVM interpretation. When the EVM executes each opcode, it must go through interpreter dispatch, stack operations, memory access, gas accounting, and other steps. These processes can further amplify costs in zero-knowledge proving or verifiable computation. RISC-V is closer to the direct execution of compiled code, and therefore can reduce the execution and proving costs of ordinary on-chain logic.

However, RISC-V cannot eliminate all performance bottlenecks. For computation-intensive tasks such as signature verification, hash computation, and large modular arithmetic, future systems will still require specialized optimization mechanisms similar to EVM precompiles. In other words, even if the execution environment shifts to RISC-V, these cryptographic primitives will not be executed solely instruction by instruction through ordinary instructions. Otherwise, the prover time will still be dominated by these heavy computational tasks.

Therefore, the key to RISC-V empowering blockchain scalability is to free ordinary business logic from EVM interpretation and reduce costs through an execution model closer to low-level machine execution.

\subsection{Execution Cost}
The advantage of RISC-V zk-VMs is not that they make all execution logic simple, but that they place part of the complex computation within a more unified and verifiable execution framework.
In practice, RISC-V zk-VMs often rely on many precompiles to handle computationally intensive operations, such as cryptographic hashing, signature verification, and elliptic curve operations. This can significantly improve performance, but it also introduces new maintenance costs. Once the blockchain base layer changes its cryptographic primitives, developers must not only modify the software implementation, but also rewrite, optimize, and audit the corresponding proof circuits. As a result, system upgrades become more complex.

For computation-heavy tasks such as state tree computation, using zk-friendly precompiles, such as Poseidon, can indeed improve performance and reduce part of the maintenance burden for execution clients. However, for logic such as deserialization, gas metering, and protocol validity checks, the challenge is not merely computational efficiency. These components involve protocol rules, data formats, and historical compatibility, making them difficult to address fully through precompiles alone.

\subsection{Extensibility and Cryptographic Acceleration}\label{sub:Extensibility}

Blockchain security relies heavily on cryptographic primitives. For example, hash functions, digital signatures, symmetric encryption, zero-knowledge proofs, and post-quantum cryptography are all relevant to blockchain execution and verification. 
These operations appear frequently in on-chain validation, off-chain proving, cross-chain communication, and device authentication.

A systematic review of hardware implementations of elliptic curve cryptography for blockchain (e.g., secp256k1 and BLS signatures) confirms substantial FPGA-based acceleration potential for the trust kernel, though RISC-V-specific ECC work remains an identified gap~\cite{boubakri2025survey}. 
%
More recent evidence suggests that RISC-V-based cryptographic acceleration can be further extended to post-quantum schemes selected by NIST. On a RISC-V SoC FPGA platform, hardware-software co-design for Kyber and Dilithium combines customized arithmetic acceleration with software optimization and reports overall performance gains of about 3~$\sim $5~$\times$, indicating that post-quantum support is becoming a practical extension of RISC-V cryptographic acceleration research~\cite{wang2024optimized,nist2024fipsapproved}.

Researchers have confirmed that RISC-V's open ISA and custom extension architecture deliver dramatic performance gains for blockchain cryptographic primitives. For example, SHA-RV completes a SHA block in 257 cycles, improving over related RISC-V designs by 9.7 to 134.9$ times$, while reducing logic resources by 89.4\% in flip-flops and 85.2\% in lookup tables~\cite{kieu2024trusted}. Crypto-RV corroborates this: a RISC-V co-processor that unifies support for SHA-256, SHA-512, SM3, SHA3-256, and AES-128 in a single 64-bit datapath, achieving 165$\times$ to 1,061~$\times$ speedup over baseline RISC-V cores at 160 MHz with 0.851 W dynamic power.

Verifiable state transitions of blockchain become more explicit in zero-knowledge proofs, fraud proofs, and light-client verification. In these settings, computation should not only be executed, but also proven or challenged. RISC-V reserves space for standard and custom extensions, which enables dedicated instructions, coprocessors, and hardware accelerators for blockchain-oriented cryptographic workloads.

Furthermore, studies on RISC-V-based blockchain hardware have mainly followed a hardware/software co-design paradigm, in which a general-purpose RISC-V core is coupled with dedicated hardware modules to accelerate blockchain-related security primitives. Specifically, Xu \emph{et al.}  \cite{xu2025risc} develop a dual-core RISC-V SoC for IoT edge blockchain deployment, where trusted execution support and a dedicated security coprocessor were integrated to accelerate key on-chain processing functions such as random number generation, SHA-256 hashing, RSA-based encryption/signature, and secure key storage. In parallel, Amirabadi et al. \cite{amirabadi2025towards} explored a RISC-V-centric optimization framework that maps high-frequency blockchain cryptographic workloads, including SHA-256, ECDSA, AES, and Keccak, onto customized hardware accelerators and instruction-level optimizations, while further incorporating lightweight authentication, identity management, and smart-contract support for embedded blockchain-enabled IoT systems.

\subsection{The Operational Mechanics of eBIM}

\begin{figure}
    \centering
    \includegraphics[width=0.48\textwidth]{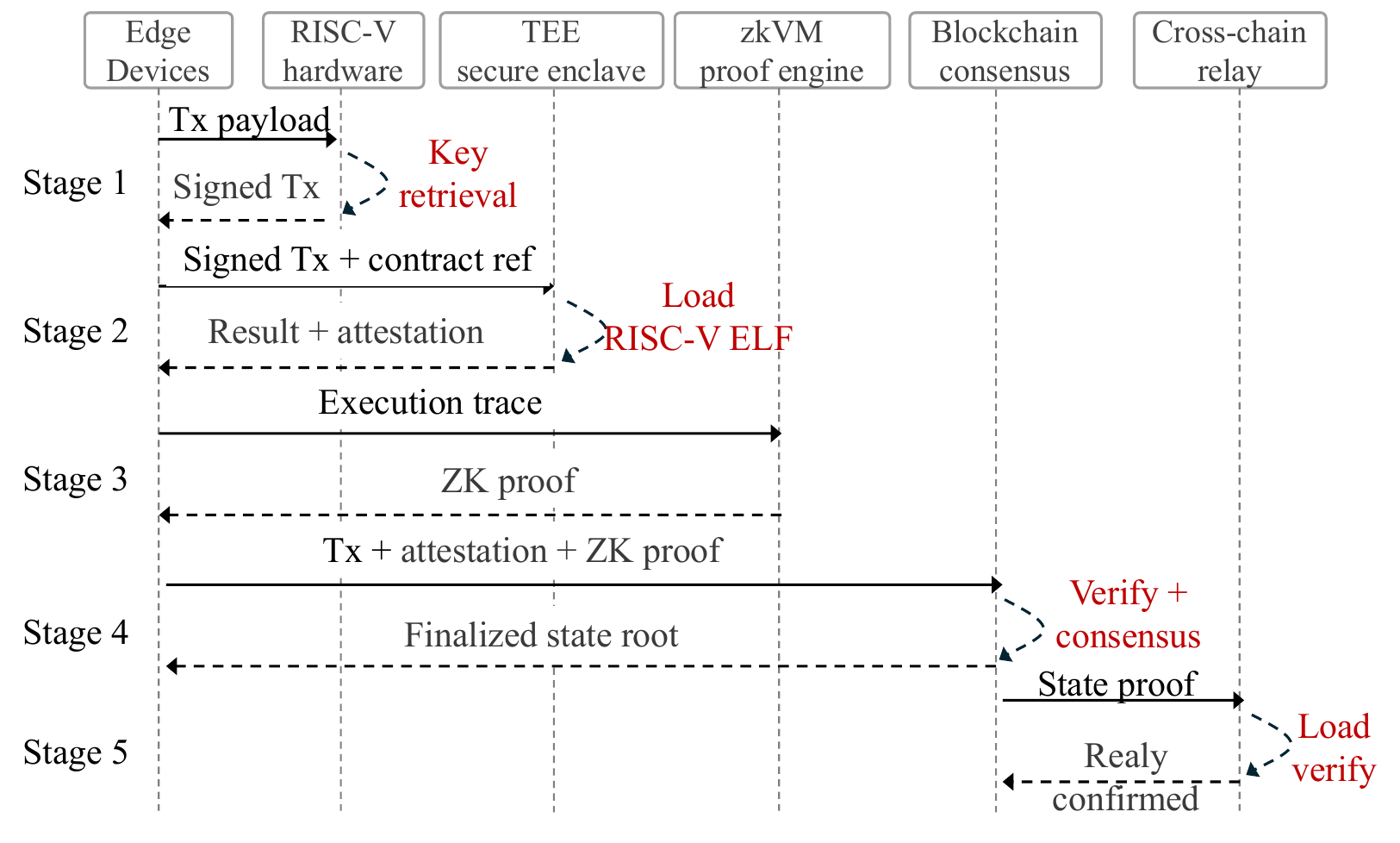}
    \caption{ \rev{Conceptual five-stage trust path for a transaction in an eBIM-enabled \new{blockchain} system. The workflow illustrates a reference composition of existing mechanisms rather than an experimentally implemented end-to-end prototype.}}
    \label{fig:executionProcess}
\end{figure}

The co-matching analysis in Subsections~\ref{sub:Openness} -- \ref{sub:Extensibility} explains why RISC-V is architecturally compatible with blockchain. Hence, this section answers how eBIM operationalizes this compatibility as a concrete transaction processing pipeline (\textbf{Answer to RQ2}). \new{Fig.\ref{fig:executionProcess} presents a conceptual reference workflow rather than an implemented end-to-end eBIM prototype. It synthesizes mechanisms that have individually been demonstrated in existing RISC-V processors, TEEs, cryptographic accelerators, zkVMs, and blockchain systems. Therefore, the five stages describe how these components may be composed to form a hardware-anchored transaction trust path, rather than implying that the complete pipeline has already been experimentally validated.}

\textbf{Stage 1: Device-level initiation and identity binding}. A transaction originates at an edge device (e.g., an IoT sensor or gateway node) equipped with an eBIM-based device. The RISC-V general-purpose CPU initializes the execution context and retrieves the device's hardware-bound private key from secure key storage. The key never leaves the chip boundary. A digital signature over the transaction payload is generated using the on-chip cryptographic accelerator (e.g., ECDSA or ML-DSA for post-quantum scenarios), producing a signed transaction object.

\textbf{Stage 2: TEE-based contract pre-execution.} The signed transaction is passed to the RISC-V SoC's TEE. Within the TEE, the relevant smart contract logic, compiled to a RISC-V ELF binary, is loaded and executed in isolation from the host operating system. Memory protection units enforce strict boundary separation. The TEE produces an execution result alongside an attestation report, cryptographically binding the output to the specific hardware and software configuration.

\textbf{Stage 3: Cryptographic proof generation.}
For transactions that require verifiable computation (e.g., ZK-Rollup or DePIN contribution proofs), the zkVM layer generates a zero-knowledge proof of the RISC-V execution trace. The cryptographic accelerator handles the computationally intensive arithmetic (hash chains, elliptic-curve operations, NTT for post-quantum schemes). \new{Dedicated accelerators may reduce the cost of computationally intensive cryptographic kernels involved in proving, such as hashing, elliptic-curve operations, or polynomial arithmetic. However, the resulting improvement in end-to-end zkVM proof-generation latency is implementation-dependent and has not yet been experimentally established for the complete eBIM pipeline.}

\textbf{Stage 4: Consensus submission and validation.} The signed transaction, TEE attestation report, and optional ZK proof are broadcast to the blockchain P2P network. Validator nodes, which may themselves run eBIM chips, independently verify: a) the cryptographic signature, b) the TEE attestation against a trusted hardware registry, and c) the ZK proof if present. Upon achieving consensus (PoS/DPoS/PBFT), the transaction is finalized, and the resulting state transition is committed to the on-chain state database.

\textbf{Stage 5: State update and cross-chain relay.} The confirmed state update is written to on-chain storage. For cross-chain transactions, a lightweight proof of the finalized state is generated and relayed to the destination chain via an interoperability protocol. eBIM chips on relay nodes perform signature verification and proof validation locally, eliminating dependence on centralized bridge operators.

\rev{This conceptual five-stage pipeline illustrates how eBIM could provide} a complete hardware-anchored trust infrastructure in which every stage of transaction processing, from origination to finalization, is grounded in verifiable, hardware-level guarantees.

\new{From a system perspective, the conceptual critical-path latency can be decomposed into device-side cryptographic processing, TEE execution and attestation, optional proof generation, network propagation and verification, consensus finalization, and optional cross-chain relay. TEE transitions and remote attestation introduce additional latency and trust-management overhead, while ZK proof generation can be substantially more expensive than ordinary transaction execution. These costs may be mitigated through batching, cached attestation evidence, asynchronous proving, and specialized accelerators, but a rigorous end-to-end evaluation remains necessary future work.}

\section{RISC-V Empowers the Development of Blockchain}\label{sec:Performance}
Blockchain platforms traditionally execute smart contracts on specialized virtual machines with bespoke instruction sets, such as the EVM.
Table~\ref{tab:arch} highlights why RISC-V is particularly suitable for blockchain workloads compared with alternative architectures. Traditional architectures such as x86 and ARM have mature toolchains and strong industrial ecosystems, but they are proprietary and therefore still require trust in specific vendors. FPGA and ASIC solutions offer high power efficiency and strong customization capabilities, yet they are less flexible within the software ecosystem and are not directly designed for smart contract execution. WASM has already been used in production smart contract platforms and benefits from an open standard. Still, it lacks native custom instruction extensions and is less suitable for zkVM-oriented execution. In contrast, RISC-V combines several advantages that are highly aligned with blockchain requirements: it is open and royalty-free, supports custom extensions, has emerging open TEE ecosystems such as Keystone and Penglai, shows high suitability for zkVMs, can support both production and proposed smart contract execution, and does not require vendor trust. Therefore, the table demonstrates that RISC-V is not merely another processor architecture but a more transparent, extensible, and verification-friendly foundation for next-generation blockchain infrastructure.

At the software level, RISC-V empowers blockchain systems by providing a more regular execution target, a more proof-friendly semantic layer, and a more reusable toolchain ecosystem, helping relieve several challenges in current blockchain execution environments.

\paragraph{Execution efficiency.}
Many blockchain systems rely on virtual machine-based execution, such as Ethereum, Solana, and Move. In Ethereum, the EVM provides deterministic smart contract execution, but its 256-bit stack-machine design creates optimization barriers. This execution model does not naturally align with the register-based organization of modern processors. RISC-V provides a register-based and compiler-friendly instruction abstraction. Hence, contract logic, execution clients, or off-chain programs can be compiled into RISC-V binaries. The resulting execution path is closer to conventional processor semantics.

\paragraph{Proof-oriented execution.}
Verifiable computation has become central to modern blockchain scaling. ZK-Rollups, zkVMs, fraud proofs, and light-client protocols all require computation to be proven or challenged. It is expensive to prove EVM-style execution directly. The stack-machine semantics, complex state access patterns, and cryptographic operations increase the cost of arithmetization. 
They may also enlarge execution traces and circuit constraints. RISC-V provides a more regular instruction-level target. A zkVM can prove the execution trace of a RISC-V program rather than designing a separate circuit for each high-level instruction.

\paragraph{Deterministic execution and isolation.}
Deterministic execution is a fundamental requirement for blockchain consensus, as blockchain nodes operate as replicas of a distributed state machine. Once a consensus protocol establishes a common ordering of transactions, all honest validating nodes must independently execute the same transactions and obtain identical state transitions, regardless of differences in their underlying hardware, operating systems, or execution environments. The EVM represents an established approach to consensus-oriented virtual machine design. The EVM specifies a stack-based instruction set, well-defined state transition semantics, and a gas-based resource accounting mechanism, providing a common execution model for Ethereum-compatible blockchain clients. Its execution semantics explicitly define arithmetic operations, storage access, control flow, exceptional termination, and interactions with the blockchain state. These specifications enable independent client implementations to reproduce the same execution results for a given transaction and pre-state under the same protocol rules. 

WebAssembly (WASM) provides an alternative execution substrate characterized by a portable, structured instruction format, sandboxed memory model, and support for multiple high-level programming languages through existing compiler toolchains. These properties make WASM attractive for blockchain platforms seeking greater programming-language flexibility and opportunities for runtime optimization. However, WASM was designed as a general-purpose execution target rather than specifically for replicated state-machine consensus.

A RISC-V-based execution environment offers another architectural approach by using a standardized instruction-set architecture as the foundation of smart contract execution. Unlike the stack-based EVM, RISC-V adopts a register-based load/store architecture with explicitly specified instruction encodings and architectural state transitions. Its modular design enables a blockchain runtime to select a constrained subset of instructions, such as an integer-oriented RV32 or RV64 profile, while excluding or tightly specifying optional extensions whose behaviors complicate deterministic execution. A consensus-oriented RISC-V virtual machine can additionally define deterministic memory initialization, address-space boundaries, exception handling, instruction counting, and resource exhaustion semantics.

\paragraph{Software ecosystem integration.}
RISC-V also provides a common semantic target across different blockchain software layers.  At the smart contract layer, RISC-V can serve as the execution substrate of contract virtual machines, as shown by systems such as CKB-VM and PolkaVM. 
At the proving layer, it can serve as the CPU semantic layer of zkVMs.  At the off-chain layer, it can support verifiable computation and dispute resolution. 
These roles are different, but they share the same instruction-level foundation. 
This makes RISC-V useful not only as a contract execution engine, but also as a unifying software abstraction for blockchain execution, proving, and verification.

Current research on blockchain-oriented RISC-V platforms still mainly focuses on traditional cryptographic schemes. However, the transition to post-quantum cryptography is becoming increasingly important for blockchain identity systems, digital signatures, and cross-chain authentication. NIST finalised its first PQC standards in 2024 (ML-KEM/Kyber, ML-DSA/Dilithium,
SLH-DSA/SPHINCS+). Blockchain systems using ECDSA and BLS12-381 face a long-term migration requirement. RISC-V's extensible ISA makes it an ideal PQC acceleration
platform~\cite{lee2023,dimatteo2025}, but the current evidence base addresses hardware implementations without blockchain integration evaluation. Future work should characterise the performance implications of PQC migration for transaction throughput and
consensus latency.

\begin{table*}[htbp]
\centering
\caption{Comparative analysis of RISC-V and alternative
architectures for blockchain workloads.}
\label{tab:arch}

\rev{
\begingroup
\small
\renewcommand{\arraystretch}{1.35}
\setlength{\tabcolsep}{4pt}

\begin{tabularx}{\textwidth}{
    >{\hsize=1.55\hsize}L
    >{\hsize=0.85\hsize}C
    >{\hsize=0.85\hsize}C
    >{\hsize=1.00\hsize}C
    >{\hsize=0.85\hsize}C
    >{\hsize=0.90\hsize}C
}
\toprule
\textbf{Dimension} &
\textbf{x86} &
\textbf{ARM} &
\textbf{FPGA/ASIC} &
\textbf{WASM} &
\textbf{RISC-V} \\
\midrule

ISA/specification openness &
Proprietary ISA\cite{hennessy2019new} &
Proprietary ISA\cite{hennessy2019new} &
Implementation-specific\cite{javaid2022blockchain} &
Open standard (e.g., W3C)\cite{WebAssembly2024} &
Open ISA\cite{riscv-ratified-specs,riscv-unpriv-2024,wang2024optimized} \\

\addlinespace[3pt]

User-defined ISA extensions &
Vendor-controlled\cite{hennessy2019new} &
Vendor-controlled\cite{hennessy2019new} &
Fully customizable hardware\cite{agrawal2022efficient,javaid2022blockchain} &
Not ISA-level &
Custom ISA extensions supported \cite{riscv-ratified-specs,riscv-unpriv-2024} \\

\addlinespace[3pt]

TEE examples &
SGX/TDX &
TrustZone/Arm CCA &
Design-specific\cite{subramanyan2017formal} &
Host-dependent &
Keystone, Penglai (open)\cite{boubakri2025survey,Feng2021Scalable,Lee2020Keystone} \\

\addlinespace[3pt]

Direct blockchain VM/guest-ISA use &
Mainly host execution\cite{buterin2016ethereum} &
Mainly host execution\cite{khan2021blockchain} &
Specialized acceleration\cite{javaid2022blockchain,lu2023scu} &
Production smart-contract execution &
Production use (CKB-VM), emerging use (PolkaVM)\cite{polkavm2025,Xue2019CKBVM} \\

\addlinespace[3pt]

zkVM guest-ISA adoption &
Not a mainstream guest ISA\cite{succinct2025sp1} &
Not a mainstream guest ISA\cite{succinct2025sp1} &
Specialized proof acceleration\cite{agrawal2022efficient,javaid2022blockchain} &
Some proof-oriented approaches &
Widely used in several current zkVMs\cite{succinct2025sp1} \\

\addlinespace[3pt]

Toolchain maturity &
Very high\cite{hennessy2019new} &
Very high\cite{hennessy2019new} &
Mature hardware toolchains\cite{javaid2022blockchain} &
Mature toolchain ecosystem\cite{WebAssembly2024}&
Mature GCC/LLVM support; growing ecosystem\cite{amirabadi2025towards,riscv-gnu-toolchain} \\

\addlinespace[3pt]

Hardware customization &
Limited at ISA/user level\cite{hennessy2019new} &
Vendor/license-dependent\cite{hennessy2019new} &
Very high\cite{agrawal2022efficient,javaid2022blockchain} &
N/A &
High, implementation-dependent\cite{amirabadi2025towards,waterman2014risc,wang2024optimized,xu2025risc} \\

\bottomrule
\end{tabularx}
\endgroup
}
\end{table*}

\section{The Initial Exploration: Utilization of RISC-V}\label{sec:Application}
As an open, modular, and extensible instruction set architecture, RISC-V has moved from academic research to broader engineering practice. Existing studies show that successful use cases of RISC-V mainly fall into three directions: i) hardware foundation; ii) cryptography; iii) blockchain~\cite{greengard2020will,amirabadi2025towards}. This section discusses the mainstream applications of RISC-V from these three perspectives.

\subsection{RISC-V Application Cases Hardware Foundation}
A successful application of RISC-V in hardware foundation is its role as a basic architecture for general system design and emerging computing platforms~\cite{frolov2021investigation}. For example, studies on the RISC-V vector extension (RVV) indicate that it supports a vector-length-agnostic programming model and can serve high-performance computing workloads, such as SeisSol and MiniFALL3D in earth sciences \cite{banchelli2026exploring}. This case is significant because it demonstrates that RISC-V is available not only to lightweight embedded systems but also to high-performance, vector-oriented scientific computing. In this context, RISC-V meets HPC requirements through RVV-based vector instruction support, compiler-driven auto-vectorization, and architecture-aware code optimization, which collectively improve the performance of scientific kernels while maintaining portability.

Furthermore, RISC-V is used to empower edge deep learning platforms and large language model inference platforms. Survey studies find that RISC-V is becoming an important ISA for edge deep learning because it's easy to integrate with SIMD, vector, systolic-array, and in-memory computing microarchitectures~\cite{agosta2025deep}. 
Garcia~\emph{et.al}~\cite{garcia2025inference} design a 64-core RISC-V processor with RVV, which can already support AI reasoning, such as BERT, GPT-2, Gemma, LLaMA, and DeepSeek-LLM. In this field, RISC-V is combined with applications through a multi-core architecture, RVV support, and optimized software stacks such as OpenBLAS, BLIS, and PyTorch, which are applied to improve matrix multiplication and model inference performance \cite{garcia2025inference}.

From an architectural perspective, RISC-V has been widely used as a foundation for new processor design and open hardware ecosystems. Studies discuss the value of RISC-V in terms of open ISA, processor design freedom, cross-platform capability, and ecosystem potential instead of targeting a specific application~ \cite{frolov2021investigation,greengard2020will}.

\subsection{RISC-V Application Cases in Cryptography}

Cryptography is one of the most direct and technically grounded application domains for RISC-V in blockchain-oriented systems. Blockchain security relies on a wide range of cryptographic operations, including transaction hashing, block-header processing, Merkle proof verification, digital signature generation and verification (as illustrated in Fig.\ref{fig:blockchain}), identity authentication, cross-chain message verification, and the arithmetic kernels used in zero-knowledge proof systems.
Compared with fixed proprietary instruction sets, RISC-V allows designers to introduce algorithm-specific or operator-level extensions while still retaining compatibility with the surrounding compiler and software ecosystem.

Post-quantum cryptography further strengthens the relevance of RISC-V-based cryptographic acceleration. Many current blockchain systems still rely on traditional public-key cryptographic schemes such as ECDSA or pairing-based signatures. Under the long-term quantum threat model, these schemes face migration pressure, especially for identity systems, long-lived assets, validator authentication, and cross-chain trust infrastructures. Post-quantum schemes, however, introduce new computational patterns. Lattice-based~\cite{dam2025accelerating} schemes require efficient polynomial arithmetic, number theoretic transform (NTT), inverse NTT, modular reduction, and matrix-vector operations~\cite{yan2026lattice}, while code-based schemes require different forms of finite-field and decoding-oriented computation~\cite{Gunasekara11217056}. These characteristics make post-quantum cryptography a suitable target for RISC-V custom acceleration.

Beyond lattice-based cryptography, RISC-V is also being explored for code-based post-quantum cryptography. Suresh et al.~\cite{suresh2025risc} present a RISC-V-based accelerator for post-quantum cryptography, targeting the acceleration of post-quantum cryptographic workloads through a RISC-V-centered architecture and showing that RISC-V can serve not only as a general-purpose control processor but also as the integration point for specialized PQC acceleration logic. Since code-based schemes exhibit different computational structures from lattice-based schemes, this case indicates that RISC-V-based PQC support should not be limited to a single algorithm family.

In general, RISC-V applications in cryptography can be summarized from three perspectives: i) RISC-V accelerates traditional cryptographic primitives that are already used in blockchain systems, such as hashing, encryption, and signature-related operations; ii) it can support post-quantum migration by accelerating key arithmetic kernels in schemes such as ML-KEM and other PQC candidates~\cite{dam2025accelerating,suresh2025risc}; iii) it provides an open and extensible hardware-software foundation for future blockchain cryptographic infrastructures. 

Nevertheless, the above works still evaluate cryptographic kernels or algorithm-level implementations rather than complete blockchain execution pipelines. Therefore, future research should focus on how RISC-V cryptographic acceleration affects transaction throughput, verification latency, consensus performance, energy efficiency, and post-quantum migration cost in real blockchain systems.

\begin{table*}[t]
\centering
\caption{Mapping between eBIM architectural blocks and representative existing systems/studies. 
The listed cases demonstrate individual building blocks or design principles related to eBIM, rather than an end-to-end implementation of the complete eBIM framework.}
\label{tab:ebim_mapping}
\scriptsize
\setlength{\tabcolsep}{4pt}
\renewcommand{\arraystretch}{1.25}
\new{
\begin{tabularx}{\textwidth}{
    p{2.6cm}
    p{3.2cm}
    X
    p{2.5cm}
    X}
\toprule
\textbf{eBIM architectural block} &
\textbf{Representative system/study} &
\textbf{Demonstrated role} &
\textbf{Evidence status} &
\textbf{Remaining gap toward eBIM} \\
\midrule

RISC-V smart-contract execution
&
CKB-VM~\cite{Xue2019CKBVM}; PolkaVM~\cite{polkavm2025}
&
Uses RISC-V as an instruction-level execution substrate for blockchain transactions or contract logic. 
CKB-VM executes RISC-V ELF binaries for on-chain script validation, while PolkaVM represents an emerging RISC-V-oriented smart-contract execution environment.
&
Deployed or testnet-level systems
&
Systematic comparison with EVM/WASM in terms of execution cost, security surface, compatibility, and developer experience remains limited.
\\

\midrule

Off-chain verifiable execution
&
Cartesi~\cite{cartesi-rollups-docs-1.5}
&
Uses RISC-V as the underlying ISA of an off-chain execution machine capable of running a Linux-based software environment, while preserving blockchain trust through on-chain coordination and dispute resolution.
&
Implemented framework
&
Integration with hardware-rooted trust, secure key management, and RISC-V TEE support has not yet been demonstrated as a unified eBIM system.
\\

\midrule

zkVM and proof-oriented execution
&
RISC Zero / Zeth~\cite{succinct2025sp1}; related RISC-V zkVMs
&
Uses RISC-V instruction semantics as the CPU-level execution trace to be proven. 
In Zeth, Ethereum execution software is compiled into RISC-V ELF binaries and executed inside the RISC Zero zkVM to generate validity proofs.
&
Implemented zkVM/prover systems
&
End-to-end co-design among zkVM proving, RISC-V hardware acceleration, TEE protection, and blockchain consensus remains an open problem.
\\

\midrule

Trusted execution environment
&
Keystone; Penglai; RISC-V TEE studies~\cite{kuhne2025dorami}
&
Provides enclave-based isolation, privilege separation, and trusted execution mechanisms on RISC-V platforms, supporting the protection of security-critical blockchain computation.
&
Research prototypes / open TEE ecosystems
&
Microarchitectural side channels, secure attestation, key protection, and formal analysis of TEE--zkVM composition require further investigation.
\\

\midrule

Cryptographic acceleration
&
SHA-RV~\cite{kieu2024trusted}; Crypto-RV; 
RISC-V-based ECC/PQC accelerators~\cite{suresh2025risc}
&
Accelerates blockchain-relevant cryptographic kernels such as hashing, digital signatures, symmetric cryptography, and post-quantum arithmetic through custom instructions, coprocessors, or dedicated hardware units.
&
Kernel-level FPGA/SoC evaluations
&
Reported speedups are mainly obtained at the cryptographic-kernel level; their impact on transaction throughput, consensus latency, proving latency, and system-level energy efficiency remains insufficiently evaluated.
\\

\midrule

Secure key storage and device identity
&
RISC-V blockchain/IoT SoC designs~\cite{xu2025risc}; 
hardware-centric RISC-V co-design~\cite{amirabadi2025towards}
&
Integrates secure key storage, authentication, cryptographic processing, and trusted execution support for blockchain-enabled IoT and edge devices.
&
Prototype / architecture-level studies
&
A complete device-identity lifecycle, including provisioning, remote update, revocation, recovery, and cross-domain migration, has not yet been established.
\\

\midrule

Cross-chain verification
&
Existing cross-chain and proof-verification studies
&
Motivates the use of local hardware for signature verification, state-proof validation, and zero-knowledge-proof verification on lightweight or edge relay nodes.
&
Primarily protocol- and architecture-level studies
&
A concrete RISC-V-based eBIM relay implementing an end-to-end cross-chain verification pipeline has not yet been demonstrated.
\\

\midrule

Upper-layer applications
&
DePIN~\cite{lin2024decentralized}; Verifiable AI~\cite{tangverifai,xing2025zero}
&
Illustrates application requirements for hardware-rooted device identity, verifiable computation, secure data provenance, and locally controlled trust in decentralized physical and AI infrastructures.
&
Application motivation / emerging use cases
&
Integrated deployments combining eBIM hardware, blockchain execution, verifiable computation, and application-level evaluation remain future work.
\\

\bottomrule
\end{tabularx}
}
\end{table*}
\subsection{RISC-V Applications in Blockchain}

At a high level, RISC-V can serve either as the target instruction set for smart contract execution or as a uniform computational semantics for verifiable execution. This dual role makes it particularly attractive for blockchain architectures that seek to improve execution efficiency, reduce proving overhead, and enhance long-term extensibility. We further clarify that these cases (Table~\ref{tab:ebim_mapping}) validate individual building blocks or design principles, rather than demonstrating a complete eBIM system.
  
\subsubsection{CKB: RISC-V for on-chain rule execution}

The Nervos Network is a multi-layered blockchain ecosystem whose Layer 1 protocol, the Common Knowledge Base (CKB), leverages RISC-V through its virtual machine, CKB-VM. CKB-VM is responsible for verifying smart contracts and transaction scripts~\cite{Xue2019CKBVM}.

To improve blockchain scalability, Nervos moves part of the computation or high-frequency interaction to upper-layer environments and organizes the system as a layered architecture. In this design, the underlying CKB stores only those states that must be globally confirmed, durably preserved, and treated as final settlement, whereas high-frequency, flexible, and application-specific computation is delegated to the upper generation layer. 

Within the layered architecture, the common knowledge layer, namely the CKB main chain, serves as the trusted base of the system. It is responsible for providing security and decentralization guarantees, maintaining a globally consistent state, supporting state verifiability, and enabling final settlement. The generation layer, by contrast, encompasses a variety of applications and scaling solutions, including clients, services, channels, and even other blockchains. 

In this framework, RISC-V serves as the instruction-semantic foundation of CKB’s on-chain rule-execution engine. Developers can write \textit{Lock Scripts} and \textit{Type Scripts} in high-level languages such as \textit{C} or \textit{Rust} and compile them into RISC-V ELF binaries. These binaries are stored in on-chain code Cells, while the script field of an ordinary \textit{Cell} references the code through fields such as `code\_hash`, `hash\_type`, and `args`. During transaction validation, a CKB node locates the relevant code Cell according to the Script attached to the input Cell, extracts the corresponding ELF binary, and loads it into CKB-VM for execution. 


The generality of CKB arises not from RISC-V alone, but from the combination of the Cell model and the RISC-V VM. The Cell model determines what kinds of state can be represented on-chain, whereas RISC-V determines how such state can be validated and constrained in a flexible programmatic manner. Openness is reflected not only in the adoption of the open and standardized RISC-V ISA, but also in the ability of developers to use multiple high-level languages and mature toolchains, as well as in the fact that the chain does not hard-code validation logic, signature schemes, or cryptographic primitives. However, the lower-level character of CKB lies in the fact that its execution environment is built on RISC-V instruction semantics that are closer to those of a real CPU than conventional contract-specific virtual machines.

\subsubsection{Cartesi: RISC-V for off-chain verifiable execution}

Cartesi's emergence was motivated by two persistent limitations of blockchain execution layers: poor scalability and limited development capability. Cartesi addresses these limitations by moving large-scale computation off-chain while preserving on-chain trust through verifiable execution and dispute-resolution mechanisms~\cite{cartesi-rollups-docs-1.5}.

Conceptually, Cartesi is a Rollups framework based on on-chain coordination, off-chain execution, and on-chain verification only when disputes arise. Its architecture separates the system into an on-chain base layer and an off-chain execution layer. The base layer is responsible for deploying dApp contracts, receiving inputs, managing assets, and handling state claims and dispute resolution, whereas the execution layer performs the actual application computation.

Cartesi selects RISC-V as the underlying ISA of the Cartesi machine because RISC-V simultaneously satisfies two fundamental requirements: general-purpose software execution and verifiable execution. As an open standard ISA, RISC-V supports Linux and its broader software ecosystem, allowing the Cartesi machine to run a complete Linux system and reuse standard libraries, file systems, process interfaces, and mature development tools such as GCC, Clang, and GDB.
This substantially improves both the engineering practicality and the development capability of blockchain application backends. Meanwhile, as a reduced instruction set architecture, RISC-V offers clear instruction semantics and relatively simple implementation, while still supporting modern operating-system features such as privilege levels, paged virtual memory, interrupts, and exceptions. 
Therefore, it can support sophisticated off-chain computation while also enabling disputes to be reduced to single-instruction state transitions, thereby lowering the cost of on-chain adjudication. 

\subsubsection{Zeth: RISC-V as the CPU semantic layer of zkVM}

By using the RISC-V instruction set as a virtualized, open execution target, developers can create a common VM language for zero-knowledge proofs (ZKPs), an optional yet increasingly important part of the blockchain model~\cite{sun2021survey}. ZKPs redefine what’s possible for privacy and scale on-chain by enabling auditable, verifiable compute for all parties and markets. 

In Ethereum and related Layer-2 scaling settings, verification of block execution results still typically depends on full nodes, infrastructure providers, or synchronization committees. To some extent, the scaling settings increase the system’s external trust assumptions and limit the generality of light-client verification, cross-chain interoperability, and zk-rollup proving systems. 
To reduce reliance on third-party execution nodes and make the correctness of block execution itself provable, RISC Zero publicly introduced Zeth in August 2023. Zeth is an open-source block prover for the Ethereum execution layer within the RISC Zero ecosystem, aiming to generate zero-knowledge validity proofs for Ethereum block execution, such that verifiers need neither to re-execute the entire block nor to place full trust in external execution nodes\cite{boundless-xyz2023Zeth}.

One of the most important features of Zeth is that it clearly demonstrates the role of RISC-V as the underlying CPU semantic layer of a zkVM. In Zeth, existing Rust-based Ethereum execution implementations are reused directly, compiled into RISC-V ELF binaries, and then executed step by step inside the RISC Zero zkVM according to RV32IM instruction semantics. Thus, although the object being proven is nominally the Ethereum block execution process, the semantic substrate that actually drives the proof is the instruction-level execution of RISC-V. The executor runs the RISC-V program and records the execution session, and the prover then generates a receipt based on that session to prove that the program was executed correctly on the given input and produced the claimed state root.

\subsection{Supporting the requirements of upper-layer applications}

In verifiable AI~\cite{tangverifai,xing2025zero}, users of cloud-based or edge-deployed models often observe only the final output, without being able to verify whether the result was produced by the claimed model, under the claimed parameters, and according to privacy or compliance constraints. This issue is especially critical in high-stakes domains such as medical diagnosis, finance, and autonomous systems. Recent work on lightweight cryptographic proofs of inference has shown that AI services require new mechanisms to balance inference efficiency and verifiability. 

Blockchain technology strengthens verifiable AI by creating immutable, timestamped records of AI predictions and computations that cannot be retroactively modified, enabling transparent accountability across distributed networks without requiring a central authority. Through smart contracts, multiple independent parties can automatically validate AI outputs against predefined criteria, while advanced cryptographic techniques like zero-knowledge proofs allow systems to demonstrate correct computation without exposing sensitive model weights or training data. 

Verifiable AI on blockchain requires computationally efficient systems that can execute or have their results verified on-chain while minimizing costs, along with cryptographic soundness through collision-resistant hashing, digital signatures, and zero-knowledge proofs. The system must ensure deterministic and reproducible outputs tied to specific model versions, with complete data integrity and provenance tracking through immutable records of inputs, metadata, and audit trails.

DePIN~\cite{lin2024decentralized} further extends the trust boundary of blockchains from digital assets to physical infrastructure. In DePIN systems, wireless devices, sensors, storage nodes, GPU clusters, energy systems, and edge computing nodes may contribute real-world resources in exchange for blockchain-based incentives. The key challenge is that on-chain reward allocation depends on off-chain physical behavior. Without trustworthy hardware, verifiable execution, and secure device identity, DePIN systems may be vulnerable to fake contributions, forged measurements, Sybil attacks, and hardware-level manipulation. Therefore, DePIN requires blockchains to function not merely as payment layers, but as coordination and verification layers for real-world infrastructure.


\textbf{Answer to RQ3.} By anchoring cryptographic operations, key management, and critical smart contract logic within a local hardware-based security domain, eBIM enables each blockchain node or device to possess autonomous trust capabilities without relying on the continuous availability of external service providers. In this paradigm, the device itself becomes an integral component of the underlying infrastructure. Trust is no longer externally provisioned or leased, but is instead intrinsically embedded within the hardware. This design is highly consistent with the core vision of DePIN: physical devices equipped with \rev{eBIM-enabled devices could potentially perform} identity authentication, data signing, and contribution verification, while the trustworthiness of on-chain states is guaranteed by hardware rather than by cloud service providers.

eBIM adopts RISC-V as its foundational architecture. The openness of RISC-V fundamentally avoids instruction-set-level licensing barriers and hidden dependencies. At the software layer, hardware acceleration interfaces for cryptographic algorithms and embedded smart contract execution environments can both be implemented based on open standards, without being bound to any specific blockchain or platform. As a result, organizations can flexibly migrate across different blockchains and application scenarios while maintaining the same underlying hardware infrastructure, thereby significantly reducing migration costs.

The customization capability of eBIM is particularly reflected at the hardware design stage. By designing dedicated acceleration units for SHA-256, elliptic curve operations, and post-quantum cryptographic algorithms, system performance is no longer constrained by the scheduling policies of general-purpose cloud computing resources, but is instead directly determined by the chip architecture itself. For scenarios with specialized performance requirements, such as high-frequency transaction signature verification or lightweight node operation on resource-constrained IoT devices, eBIM can achieve hardware-level, application-specific optimization that conventional Blockchain-as-a-Service platforms are unable to provide.

eBIM confines key operations and the execution of sensitive smart contract logic within an isolated security domain of the local chip, enabling trusted verification to be completed without requiring data or computation to leave the device. This allows organizations to participate in blockchain networks while satisfying local data sovereignty requirements, shifting regulatory compliance control back to the data owner rather than relying solely on the compliance commitments of service providers. Furthermore, hardware support for post-quantum cryptographic algorithms enhances the long-term security and compliance of key management systems, which is particularly important for highly regulated sectors such as finance and healthcare.

\section{Challenges and Future Directions}\label{sec:Challenges}

Current research on RISC-V for blockchain has gradually expanded from cryptographic primitive acceleration to edge SoC design, trusted execution, and collaborative architectures for blockchain-enabled IoT systems. Existing progress is meaningful, but it is still fragmented and has not yet formed a complete and mature technical system. The main challenges and future directions can be summarized as follows.

\subsection{A unified RISC-V cryptographic extension standard for blockchain}

Many existing studies have proposed RISC-V extensions or coprocessor designs for common blockchain-related cryptographic primitives, such as SHA-256, SHA-3, AES, ECC, and RSA. However, most of these efforts are designed for a single algorithm, a single platform, or a specific application scenario. As a result, there are significant differences across existing works in instruction design, data path organization, buffering mechanisms, and hardware-software interfaces. A unified RISC-V cryptographic extension standard specifically for blockchain workloads has not yet been established.

In the future, more effort is needed to develop standardized RISC-V crypto extensions for blockchain applications. Such a standard should support key operations, including hashing, digital signature generation and verification, symmetric encryption, identity authentication, and on-chain verification, under a unified software toolchain and hardware ecosystem, helping reduce integration cost, improve portability, and promote the practical deployment of blockchain-oriented RISC-V platforms.

\subsection{Smart Contract and
 Off-chain}
\label{sec:academic_lag}

Three major blockchain platforms have adopted RISC-V as their smart contract execution layer, yet peer-reviewed academic evaluation is essentially absent, a profile
characteristic of a field where industrial deployment has outpaced academic investigation by approximately two to three years. Priority research questions: 1) formal performance characterisation of PolkaVM versus EVM and WASM~\cite{WebAssembly2024} across DeFi workloads; 2) security analysis of the RISC-V contract execution surface relative to EVM; 3) developer experience assessment for Solidity-to-RISC-V compilation pipelines; 4) comparative analysis of RISC-V versus WASM for blockchain VM construction.

A bug in a zkVM constraint system could permit generation of valid proofs for incorrect computations, a catastrophic safety failure for any system relying on zkVM correctness. Relevant foundational work includes the CertikOS/SeKVM line on formally verified hypervisors, the EasyCrypt and Jasmin frameworks for formally verified cryptographic implementations, and the SAIL model providing machine-readable formal semantics of the RISC-V ISA in Isabelle, a critical foundation for reasoning about RISC-V circuit correctness. The finding that standard LLVM~\cite{balasubramanian2024designing} optimisations transfer poorly to zkVM proof constraint minimisation motivates zkVM-specific compiler verification frameworks, which do not yet exist.

\subsection{TEEs with RISC-V}
\label{sec:sidechannel}
The TEE-protected proof generation pattern has not been formally analysed for security. Key open questions: What are the trust assumptions when a TEE protects a zkVM proving process? Does TEE compromise invalidate the zero-knowledge property of the resulting proof? These questions sit at the intersection of TEE security analysis and ZK proof theory and constitute a natural research frontier.

Cache-timing side-channel attacks have been demonstrated against Keystone-based RISC-V TEE implementations~\cite{kuhne2025dorami}. These attacks exploit shared cache state to infer enclave execution patterns without violating the formal isolation guarantee. For blockchain validator key management where even partial key leakage is catastrophic, side-channel resistance is a first-class requirement. Current RISC-V TEE frameworks address logical isolation but provide limited mitigations for micro-architectural side channels. Dedicated cache partitioning, constant-time execution enforcement, and side-channel-resistant memory protection extensions are needed.
\subsection{Efficient cross-chain verification}

Existing studies have already started to discuss cross-chain communication and interoperability. However, most of them remain at the protocol or architecture level. In practice, cross-chain verification usually involves inter-chain message proof verification, state proof validation, signature checking, and even zero-knowledge proof verification. These operations are computationally expensive, especially for edge devices and light nodes.

With the rapid development of verifiable execution and zkVM technologies, RISC-V has the potential to become an important hardware foundation for lightweight cross-chain verification nodes and proof-verification terminals.

\subsection{Post-quantum security}
Post-quantum cryptography strengthens the case for RISC-V-based hardware acceleration. NIST has standardized several major post-quantum cryptographic schemes, including ML-KEM, ML-DSA, and SLH-DSA, signaling the beginning of practical post-quantum migration. For blockchains, this transition is especially consequential because public keys, signatures, long-lived assets, and historical transactions may remain exposed for extended periods. Post-quantum schemes often involve computationally intensive operations such as polynomial multiplication, number-theoretic transforms, sampling, hashing, and matrix-vector arithmetic. RISC-V’s extensible design enables hardware/software co-design for such workloads. Existing research on RISC-V-based accelerators for post-quantum cryptography, including accelerators for CRYSTALS-Dilithium, Falcon, and Classic McEliece, shows that RISC-V can support algorithm-specific instruction extensions and coprocessors while preserving software programmability.

A major challenge is that post-quantum cryptographic algorithms are highly diverse. They include lattice-based, hash-based, and code-based schemes, and their computation patterns differ much more significantly than those of traditional cryptographic algorithms. Therefore, future RISC-V platforms should not only support current blockchain cryptographic primitives but also provide efficient and unified support for post-quantum algorithms. At the same time, such support should consider side-channel resistance, hardware area efficiency, and the constraints of edge deployment.

\subsection{Hardware acceleration: system-level co-design}

A large body of existing work focuses on accelerating a single cryptographic component, such as hashing, signature verification, or encryption. These studies have achieved clear improvements in performance and energy efficiency. However, real blockchain execution is not composed of isolated algorithm calls. Instead, it involves a complete system flow, including data movement, buffer scheduling, authentication procedures, pre-on-chain processing, smart contract execution, and security isolation. The post-quantum case makes this system-level requirement even clearer. Recent work on Ethereum-based blockchains estimates the integration cost of post-quantum algorithms and shows that migration should be assessed at the system level rather than treated as a simple algorithm substitution. This is consistent with RISC-V co-design results for Kyber and Dilithium, where arithmetic acceleration, hashing support, and software optimization must be considered together to obtain practical gains~\cite{juaristi2024benchmarking,wang2024optimized}.

Therefore, future work should move beyond simply adding more accelerators for more algorithms. A more important direction is system-level hardware-software co-design. This may include unified internal buffering mechanisms, dedicated DMA support for on-chain data flow, reconfigurable cryptographic execution units, and the integration of near-memory or in-memory computing into edge blockchain architectures.

\subsection{Trust identity management}

In IoT and edge environments, one of the most important values of blockchain lies in trusted identity management. However, most current studies mainly focus on identity signatures, access control, or hierarchical identity mechanisms. They have not yet fully addressed more practical electronic identity systems required by real-world devices, such as eSIM support, hardware root of trust, remote identity update, and cross-domain identity recognition.
In the future, if blockchain-enabled IoT systems are expected to support large-scale device access, RISC-V platforms must play a deeper role in the full lifecycle of device identity management. This includes identity generation, identity binding, authentication, recovery, revocation, and cross-domain migration.

Based on current research trends, the most practical application of blockchain-oriented RISC-V is not in large public-chain full nodes, but in portable devices, edge gateways, industrial controllers, and IoT terminals. However, these scenarios also face multiple challenges at the same time, including limited hardware resources, strict power budgets, heterogeneous network environments, physical exposure, and real-time requirements.

Therefore, future studies should continue to focus on lightweight design, multi-layer collaboration, and cloud-edge coordination. The central question is whether edge devices can perform blockchain-related tasks securely, efficiently, and sustainably.

\section{Conclusions}\label{sec:Conclusions}
In this study, the embedded Blockchain Infrastructure Management (eBIM) as a software-hardware collaborative paradigm for next-generation blockchain infrastructures is introduced. By leveraging the openness, modularity, extensibility, and verification-friendly design of RISC-V, eBIM provides a promising foundation for integrating cryptographic acceleration, trusted execution, key management, and smart contract support into the hardware layer. This shift enables blockchain nodes, edge devices, and decentralized infrastructures to achieve greater autonomy, verifiability, and security without excessive reliance on centralized service providers.

The proposed eBIM framework is not merely an alternative execution backend for blockchain systems, but a low-level substrate that can support deterministic execution, zkVMs, trusted computing, post-quantum cryptography, and hardware-software co-design. Representative applications, including smart contract execution, cryptographic acceleration, DePIN-oriented devices, Verifiable AI, BaaS, and Infrastructure as Code, demonstrate the broad potential of RISC-V-empowered blockchain systems. Nevertheless, eBIM remains an emerging research direction. 

\new{Therefore, the principal significance of eBIM is not the proposal of another blockchain consensus protocol or virtual machine, but the establishment of a common architectural viewpoint for moving selected blockchain trust functions from externally managed infrastructure into an open, programmable, and locally controlled hardware/software substrate. By connecting device identity, secure key management, trusted execution, cryptographic acceleration, verifiable computation, and blockchain validation within the same framework, eBIM provides a basis for studying the trust continuum from edge hardware to blockchain consensus. At present, this significance is primarily architectural: complete prototypes, quantitative end-to-end benchmarks, and deployment-scale security evaluations remain important future research directions.}

\bibliographystyle{cas-model2-names}

\bibliography{refs}


\newpage
\bio{QinglinYang}
Qinglin~Yang received his Ph. D. degree in Computer Science and Engineering from the University of Aizu, Japan, 2021. He has been a guest editor for a blockchain special issue at IEEE OJCS. He also served as a TPC chair for several machine-learning conferences. He is an assistant professor at the Cyberspace Institute of Advanced Technology/Huangpu Research School of Guangzhou University, Guangzhou University, China. His current research interests include blockchain, cybersecurity, and federated learning. 
\endbio
\bio{YuanLiu}
Yuan Liu is a professor at the Cyberspace Institute of Advanced Technology of Guangzhou University in Guangdong, China. She achieved her Ph.D degree in the School of Computer Engineering from Nanyang Technological University (NTU), Singapore, in 2014.  She was an associate professor at Northeastern University, China, from 2015 to 2022. From 2014 to 2015, she worked as a Research Fellow at the Joint NTU-UBC Research Center of Excellence in Active Living for the Elderly (LILY), NTU, Singapore. Her research interests include incentive mechanism design, federated learning, trust management, blockchain consensus protocols, blockchain-empowered artificial intelligence, and threat intelligence systems. 

\endbio

\vspace{10mm}

\bio{YaoyaoZhang}
 Yaoyao Zhang is pursuing her PhD degree with the Cyberspace Institute of Advanced Technology of Guangzhou University in Guangdong, China. Her research interests include blockchain and data trading, cybersecurity, and federated learning. 
\endbio
\vspace{15mm}

\bio{BoyaWang}
Boya~Wang received her B. Sc. degree from Guangzhou University, Guangzhou, China, in 2025. She is currently pursuing her M. Sc. degree with the Cyberspace Institute of Advanced Technology, Guangzhou University, Guangzhou, China. Her research interests include cybersecurity and blockchain.
\endbio

\vspace{20mm}

\bio{ZongjianYou}
Zongjian You received his B. Eng. degree from Fuzhou University, Fuzhou, China, in 2025. He is currently pursuing his M. Sc. degree with the Cyberspace Institute of Advanced Technology, Guangzhou University, Guangzhou, China. His research interests include cybersecurity, game theory, and blockchain.
\endbio

\vspace{20mm}

\bio{ChunmingRong}
Chunming Rong (Senior Member, IEEE) is currently the head of Data-Centered and Secure Computing (DSComputing), University of Stavanger (UiS). He has also played a vital leadership role (first as vice chair and then as chair) in both the IEEE Cloud Computing initiative and the IEEE CS STC Cloud Computing, and led its transition to the IEEE CS Technical Committee on Cloud Computing (TCCLD). He has extended his engagement with the IEEE Future Directions through his involvement with the IEEE Blockchain Initiative from 2017 to 2018. He is an executive member of Technical Consortium on High Performance Computing (TCHPC) and the chair of the STC on Blockchain in the IEEE Computer Society. He is also co-founder of two start-ups bitYoga and Dataunitor, in Norway, both of which were the recipients of the EU Seal of Excellence Award in 2018. 
\endbio

\bio{ZhihongTian}
Zhihong~Tian is currently a professor with the Cyberspace Institute of Advanced Technology, and Vice President of Guangzhou University, Guangdong Province, China. Guangdong Province Universities and Colleges Pearl River Scholar (Distinguished Professor). He is also a part-time Professor at Carlton University, Ottawa, Canada. Previously, he served in different academic and administrative positions at the Harbin Institute of Technology. 
His research has been supported in part by the National Natural Science Foundation of China, the National Key Research and Development Plan of China, the National High-tech R\&D Program of China (863 Program), and the National Basic Research Program of China (973 Program). He also served as a member, Chair, and General Chair of several international conferences. He is a Senior Member of the China Computer Federation, and a Member of IEEE. 
\endbio

\end{document}